\pdfoutput=1
\documentclass[sigconf]{acmart}

\renewcommand\footnotetextcopyrightpermission[1]{} 
\AtBeginDocument{%
  \providecommand\BibTeX{{%
    \normalfont B\kern-0.5em{\scshape i\kern-0.25em b}\kern-0.8em\TeX}}}

\setcopyright{none}

\acmConference{arxiv}
\acmBooktitle{arxiv}

\usepackage{subcaption}
\usepackage{paralist}
\usepackage[linesnumbered]{algorithm2e}
\usepackage{xcolor}

\newcommand{\sql}[1]{%
	\textbf{\texttt{#1}}%
}

\begin{document}

\title{Machine Learning-based Cardinality Estimation in DBMS on Pre-Aggregated Data}

\author{Lucas Woltmann, Claudio Hartmann, Dirk Habich, Wolfgang Lehner}
\affiliation{%
  \institution{Database Systems Group, TU Dresden}
  \city{Dresden}
  \country{Germany}}
\email{firstname.lastname@tu-dresden.de}

\renewcommand{\shortauthors}{Woltmann et al.}

\begin{abstract}
Cardinality estimation is a fundamental task in database query processing and optimization. 
As shown in recent papers, machine learning (ML)-based approaches can deliver more accurate cardinality estimations than traditional approaches.
However, a lot of example queries have to be executed during the \emph{model training phase} to learn a data-dependent ML model leading to a very time-consuming training phase.
Many of those example queries use the same base data, have the same query structure, and only differ in their predicates.
Thus, index structures appear to be an ideal optimization technique at first glance. 
However, their benefit is limited. 
To speed up this model training phase, our core idea is to determine a \emph{predicate-independent pre-aggregation} of the base data and to execute the example queries over this pre-aggregated data. 
Based on this idea, we present a specific \emph{aggregate-enabled training phase} for ML-based cardinality estimation approaches in this paper. 
As we are going to show with different workloads in our evaluation, we are able to achieve an average speedup of $63$ with our \emph{aggregate-enabled training phase}.
\end{abstract}



\maketitle

\section{Introduction}
\label{sec:Intro}

Due to the increasing amount of data managed by database systems (DBMS), query optimization is still an important challenge. The main task of query optimization is to determine an efficient execution plan for every SQL query, whereby most of the optimization techniques are cost-based~\cite{leis2015good}. 
For these techniques, cardinality estimation has a prominent position with the task to approximate the number of returned tuples for every query operator within a query execution plan~\cite{DBLP:journals/pvldb/HarmouchN17,leis2015good,DBLP:journals/pvldb/MoerkotteNS09,DBLP:conf/vldb/YoussefiW79}. 
Based on these estimations, various decisions are made by different optimization techniques such as choosing (i) the right join order~\cite{DBLP:conf/icde/FenderM11}, (ii) the right physical operator variant~\cite{DBLP:conf/vldb/RosenfeldHVM15}, (iii) the best-fitting compression scheme~\cite{DBLP:journals/tods/DammeUHHL19}, or (iv) the optimal operator placement within heterogeneous hardware~\cite{DBLP:journals/pvldb/KarnagelHL17}. 
However, to make good decisions in all cases, it is important to have cardinality estimations with high accuracy. 

As shown in recent papers~\cite{kipf2018learned, woltmann2019local, liu2015nn}, machine learning-based cardinality estimation approaches are able to meet this high accuracy requirement much better than traditional estimation approaches.
While traditional approaches such as histogram-based and frequent values methods assume uniformity and data independence for their estimation~\cite{leis2015good}, ML-based approaches assume that a sufficiently deep neural network can model the very complex data dependencies and correlations~\cite{kipf2018learned}.
For this reason, ML-based cardinality estimation approaches can, of course, give much more accurate estimations as clearly demonstrated in~\cite{kipf2018learned, woltmann2019local, liu2015nn}. 
However, the main drawback of these ML-based techniques compared to traditional approaches is the high construction cost of the \emph{data-dependent} ML-model based on the underlying supervised learning approach.
During the so-called \emph{training phase}, the task of \emph{supervised learning} is to train a model, or more specifically learn a function, that maps input to an output based on example \texttt{(input,output)} pairs. 
Thus, in the case of cardinality estimation, a lot of pairs consisting of \texttt{(query, output-cardinality)} are required during the \emph{training phase}.
To determine the right \texttt{output-cardinalities}, the queries have to be executed~\cite{kipf2018learned, woltmann2019local}, whereby the execution of those example queries can be very time consuming, especially for databases with many tables, lots of columns, and millions or billions of tuples. 
Moreover, this \emph{training phase} leads to a heavy, selective load on the database.

\textbf{Core Contribution.}
To overcome these shortcomings, we propose a novel training phase \emph{based on pre-aggregated data} for ML-based cardinality estimation approaches in this paper.
As described in~\cite{kipf2018learned, woltmann2019local}, every example query is (i) rewritten with a count aggregate to retrieve the right \texttt{output-cardinality} and (ii) executed individually. 
However, many of those example queries use the same base data, have the same query structure, and only differ in their predicates.
On the one hand, index structures on the base data are already helping to speed up the query execution.
On the other hand, their benefit is limited as we are going to show in our evaluation.
To further optimize the query execution, our core idea is to determine a \emph{predicate-independent pre-aggregation} of the base data and to execute the example queries over this pre-aggregated data. 
The advantage is that this \emph{pre-aggregation} already removes all redundancies from the base data. 
Then, the set of similar example queries has to read and process less data.
To realize this \emph{pre-aggregation}, the most common solution in DBMS is to create a \emph{data cube} for storing and computing aggregate information~\cite{gray1996cube}. 
However, this \emph{pre-aggregation} is only beneficial if the execution of the example queries on the data cube including the time for creating the data cube is faster than the execution of the example queries over the base data.
As we are going to show with different workloads for the training phase in our evaluation, we are able to achieve an average speedup of $82$ which clearly outperforms the optimization using index structures on base data. 


\textbf{Contributions in Detail and Outline.}
Our \emph{aggregate-enabled training phase} consists of two phases: (i) creation of a set of meaningful \emph{pre-aggregated data sets} using data cubes and (ii) rewrite and execute the example queries on the corresponding data cubes or the base data. 
To present our overall approach, we make the following contributions in this paper:
\begin{compactenum}
\item We start with a general overview of ML processes in DBMS in Section~\ref{sec:models}. 
In particular, we detail cardinality estimation as a case study for ML in a DB. 
We introduce \emph{global} and \emph{local models} as two representatives for ML-based cardinality estimation approaches.
In particular, we show their properties in terms of example workload complexity and conclude the need for optimization of such workloads.
\item Based on that, we introduce our general solution approach for an \emph{aggregated-enabled training phase} by \emph{pre-aggregating} the base data using the \emph{data cube} concept and executing the example queries over this pre-aggregated data. 
Moreover, we introduce a \emph{benefit criterion} to decide whether the \emph{pre-aggregation} makes sense or not.
\item In Section~\ref{sec:training}, we present our \emph{aggregate-enabled training phase} for ML-based cardinality estimation approaches in detail.
Our approach consists of two components: \texttt{Analyzer} and \texttt{Rewrite}. While the main task of the \texttt{Analyzer} component is to find and build all beneficial data cubes, the \texttt{Rewrite} component is responsible for rewriting the example queries to the constructed data cubes if possible. 
\item Then, we present experimental evaluation results for four different workloads for the training phase of ML-based cardinality estimation in Section~\ref{sec:eval}. The workloads are derived from different ML-based cardinality estimation approaches~\cite{kipf2018learned,woltmann2019local} on the IMDB data set~\cite{imdb}. 
Moreover, we compare our approach with the optimization using index structures.
\end{compactenum}
Finally, we close the paper with related work in Section~\ref{sec:related} before concluding in Section~\ref{sec:conclusion}.

\section{Machine learning Models for DBMS}
\label{sec:models}

In this section, we start with a brief description of the general process of \emph{machine learning (ML)} in the context of DBMS. 
Then, we discuss ML-based cardinality estimation for DBMS as an important case study and introduce two ML-based approaches solving this specific challenge. 
Finally, we analyze the specific query workloads for the training phases and clearly state the need for optimized database support.

\begin{figure}[t]
    \centering
    \includegraphics[trim=100 75 100 75, clip, width=0.95\columnwidth]{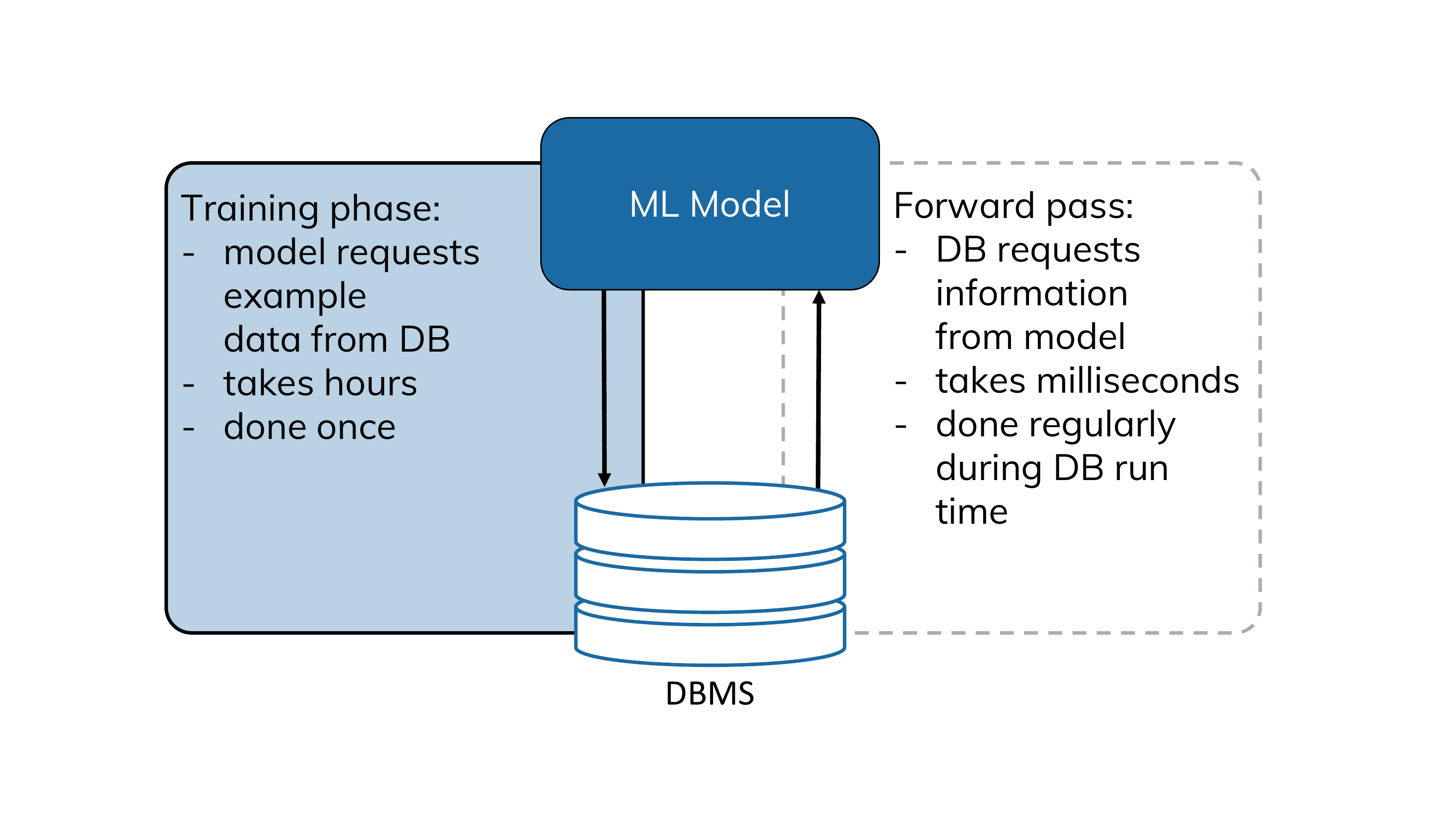}
    \caption{The general process of supervised ML in DBMS.}
    \label{fig:process}
    \vspace{-0.4cm}
\end{figure}
\begin{figure*}
    \centering
    \includegraphics[width=\textwidth]{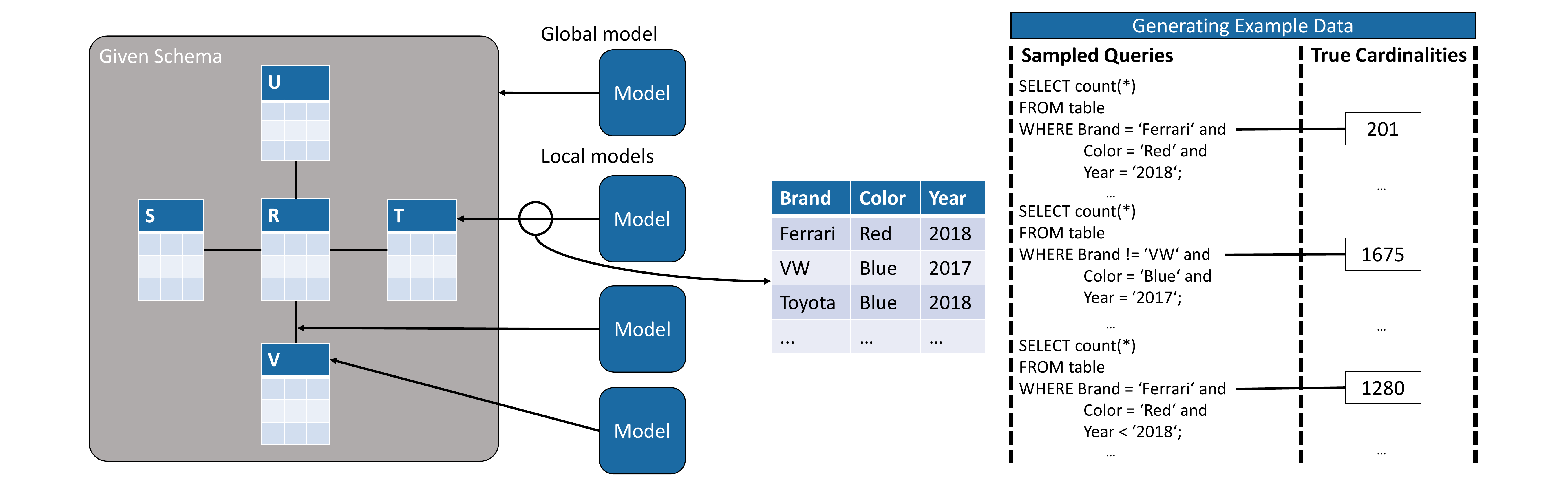}
    \caption{Overview of ML-based cardinality estimation approaches.}
    \label{fig:workload}
    \vspace{-0.4cm}
\end{figure*}

\subsection{Machine Learning Support for DBMS}

Most ML-supported techniques for DBMS are supervised learning problems.
In this category, there are amongst others: cardinality estimation~\cite{liu2015nn, kipf2018learned, woltmann2019local}, plan cost modeling~\cite{ryan2019plan, ji2019cost}, and indexing~\cite{kraska2018case}.
The proposed ML solutions for those highly relevant DBMS problems have a general process in common as shown in Figure~\ref{fig:process}.
This process is usually split into two parts: \emph{forward pass} and \emph{training phase}.
\begin{compactenum}
\item[\textbf{Forward pass:}] This pass consists of the application of the model triggered by a request of the DBMS to the ML model. 
Each request pulls some specific information from the model such as an estimated cardinality or an index position~\cite{liu2015nn,kraska2018case,kipf2018learned, woltmann2019local}. 
The execution time of each forward pass request is normally in the range of milliseconds. 
This is advantageous because forward passes occur often and regularly during the run time of the DBMS~\cite{liu2015nn,kraska2018case,kipf2018learned, woltmann2019local}.
\item[\textbf{Training phase:}] To enable the forward pass, a training phase is necessary to construct the required ML model, whereby the challenge for the model lies in the generalization from example data~\cite{liu2015nn,kraska2018case,kipf2018learned, woltmann2019local}. 
That means, the model usually requests a lot of diverse labeled example data---pairs of \texttt{(input, output)}---from the DBMS to learn the rules of the underlying problem. 
Even though the training is done once, its run time can take hours.
This is mainly caused by the generation and execution of a large number of queries against the DBMS to determine the labeled example data. 
\end{compactenum}

That means, the \emph{training phase} of ML models to support DBMS usually generates a punctual high load on the DBMS and compared to the \emph{forward pass}, this training is significantly more expensive from a database perspective.
Therefore, the training phase is a good candidate for some specific database support and each optimization will reduce (i) the time for the training phase and (ii) the punctual load on the DBMS.
Thus, database support or optimization of the training phase is a novel and interesting research field leading to an increased applicability of ML support for DBMS in the end.

\subsection{Case Study: Cardinality Estimation}
\label{sec:casestudy}
As already mentioned in the introduction, we restrict our focus to the ML-based cardinality estimation use case~\cite{liu2015nn, kipf2018learned, woltmann2019local} in this paper.
Here, each \emph{forward pass} requests an estimated cardinality for a given query from the ML model.
In the \emph{training phase}, the ML cardinality estimator model requires example queries as example data from the DB where the queries are labeled with their true cardinality resulting in pairs of (\texttt{query}, \texttt{cardinality}).
These cardinalities are retrieved from the DB by executing the queries enhanced with a count aggregate.
For that, two major approaches for user-workload-inde\-pend\-ent cardinality estimation with ML models have been proposed in recent years: \emph{global} and \emph{local models}.
Their retrieval of example queries and the training are similar, but the models differ in their focus.
The focus or \emph{model context} describes the coverage of a model in a given database schema and which cardinalities it can estimate.
On the one hand, the DBMS can query the global and local model for cardinalities in the forward pass only in this context.
On the other hand, during training example queries are only generated from that particular context.


\subsubsection*{\textbf{Global Model Approach}}
In that sense, a \emph{global model} is trained on the \emph{whole database schema} as its model context~\cite{kipf2018learned}.
They are effective in covering correlations in attributes for high-quality estimates~\cite{kipf2018learned}.
Moreover, only one single ML model is needed to cover the whole database schema.
In Figure \ref{fig:workload}, this is depicted by just one model stated and mapped to the whole schema.
However, global models have downsides like (i) the complexity of the ML model and (ii) the very expensive training phase.
Both disadvantages arise for the following reason: the single ML model handles all attributes and joins in the same way leading to a huge problem space.  
This huge problem space is directly translated to the model complexity as well as to a high number of example queries to cover all predicates and joins over the whole schema as shown~\cite{kipf2018learned}. 


\subsubsection*{\textbf{Local Model Approach}}
To overcome the shortcomings of the global model approach, the concept of \emph{local models} has been introduced~\cite{woltmann2019local}.
Local models are ML models which only cover a certain sub-part of the whole database schema as their model context.
This can be a base table or any n-way join.
Again, Figure \ref{fig:workload} details several local models each covering a different part of the schema.
As they focus on a small part of the schema, there are many advantages compared to global models.
First of all, local models produce high-quality estimates, like global models as shown in~\cite{woltmann2019local}.
However, because they cover a smaller problem space in different combinations of predicates and joins, their model complexity is much smaller.
The lower complexity comes from a more focused or localized problem solving.
A local model has to generalize a smaller problem than the global, i.e. the cardinality estimate of a sub-part of a schema and not the whole schema at once.
The lower complexity leads to faster example query sampling and training because the easier problem requires fewer queries during training.
A major disadvantage of local models is the high number of models needed to cover a whole database schema. 
That means, for each sub-part of the schema, a separate local ML model needs to be constructed. 
This also requires additional queries to be generated because each local model requests the same amount of example queries.
However, these queries are less complex because there are fewer combinations of predicates and tables in a local context.

\begin{figure*}
    \centering
    \begin{subfigure}[b]{0.33\textwidth}
        \includegraphics[width=0.8\textwidth]{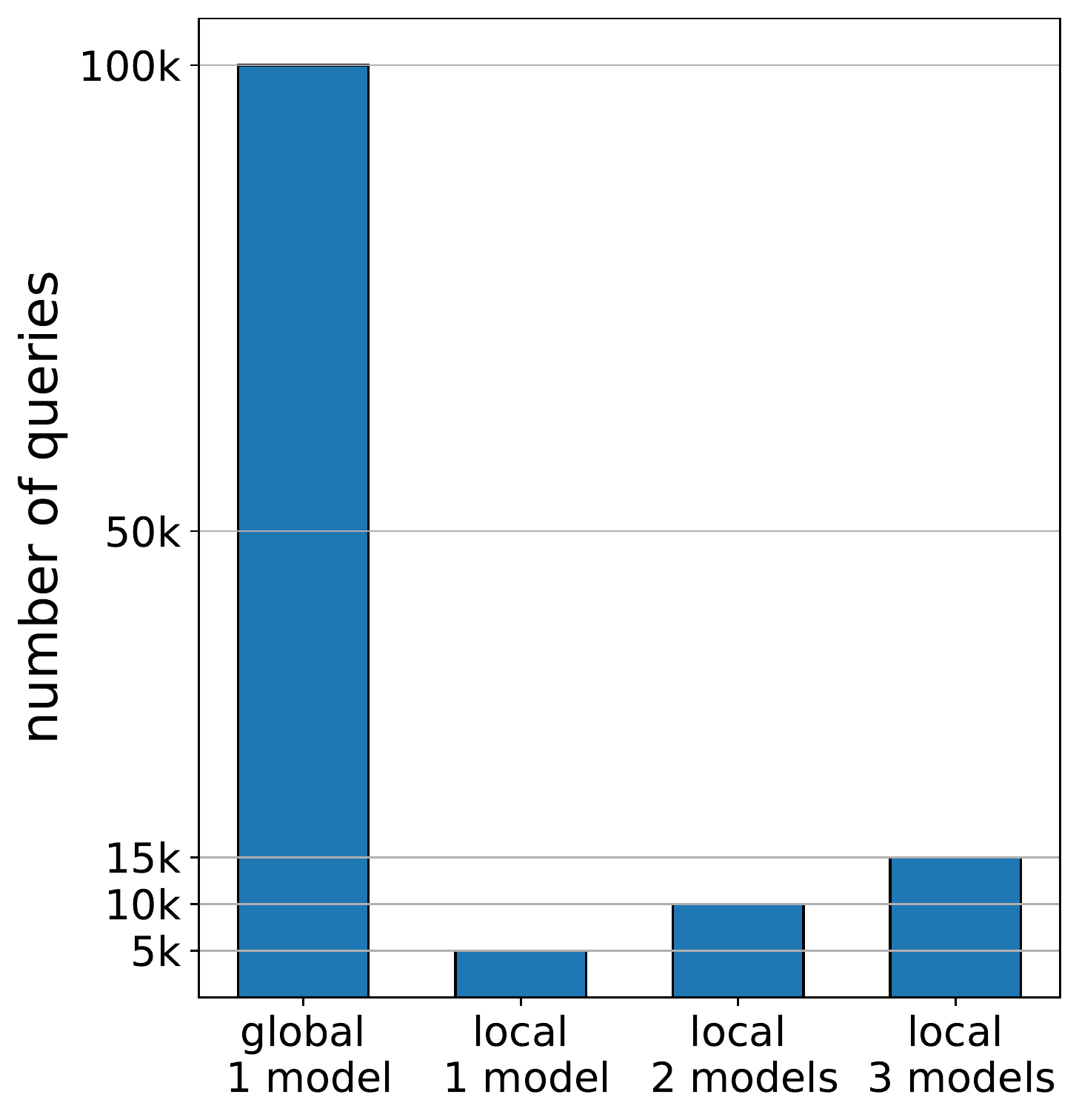}
        \caption{Number of queries.}
        \label{fig:queries}
    \end{subfigure}
    \begin{subfigure}[b]{0.33\textwidth}
        \includegraphics[width=0.8\textwidth]{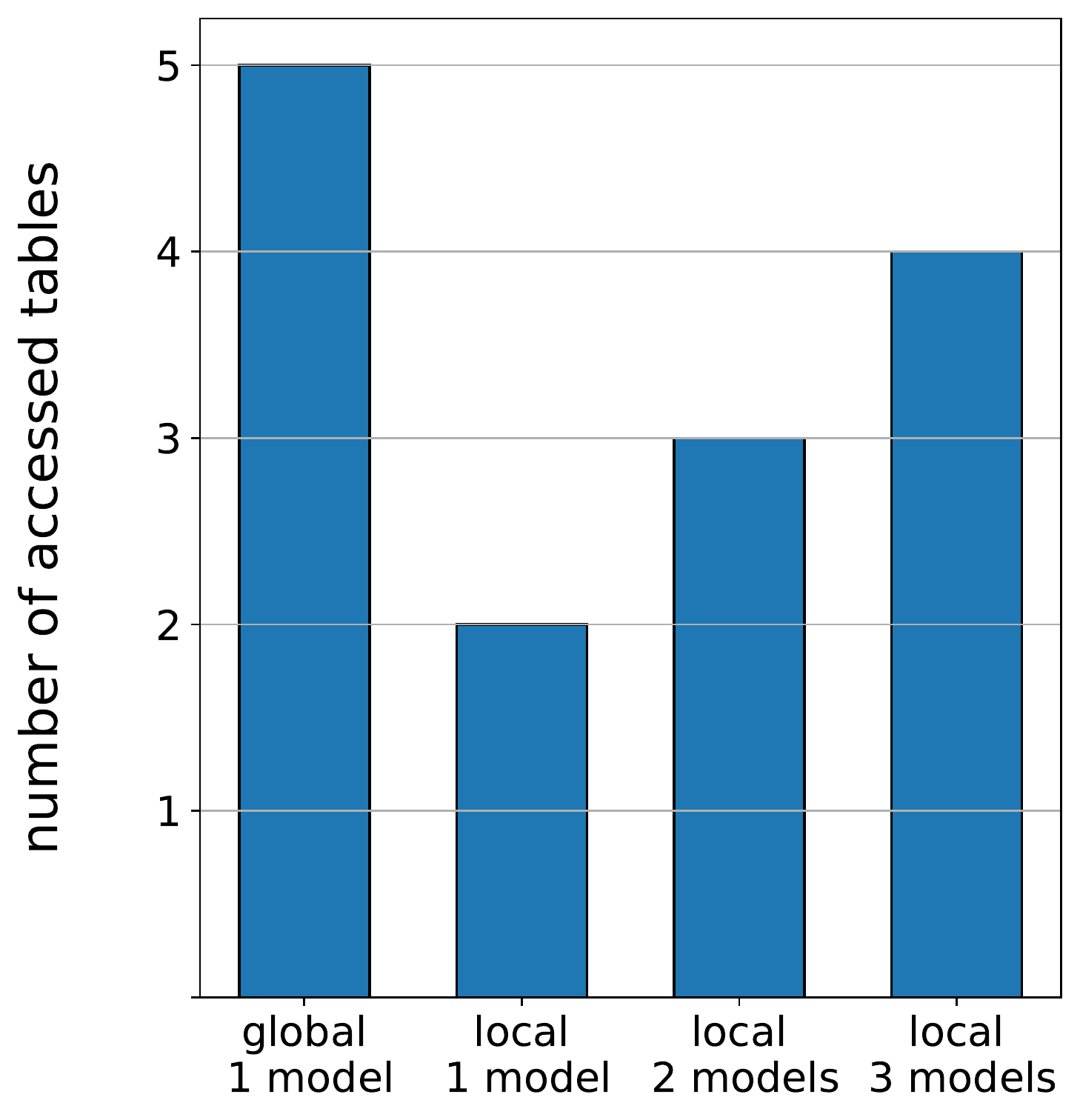}
        \caption{Number of accessed tables.}
        \label{fig:tables}
    \end{subfigure}
    \begin{subfigure}[b]{0.33\textwidth}
        \includegraphics[width=0.8\textwidth]{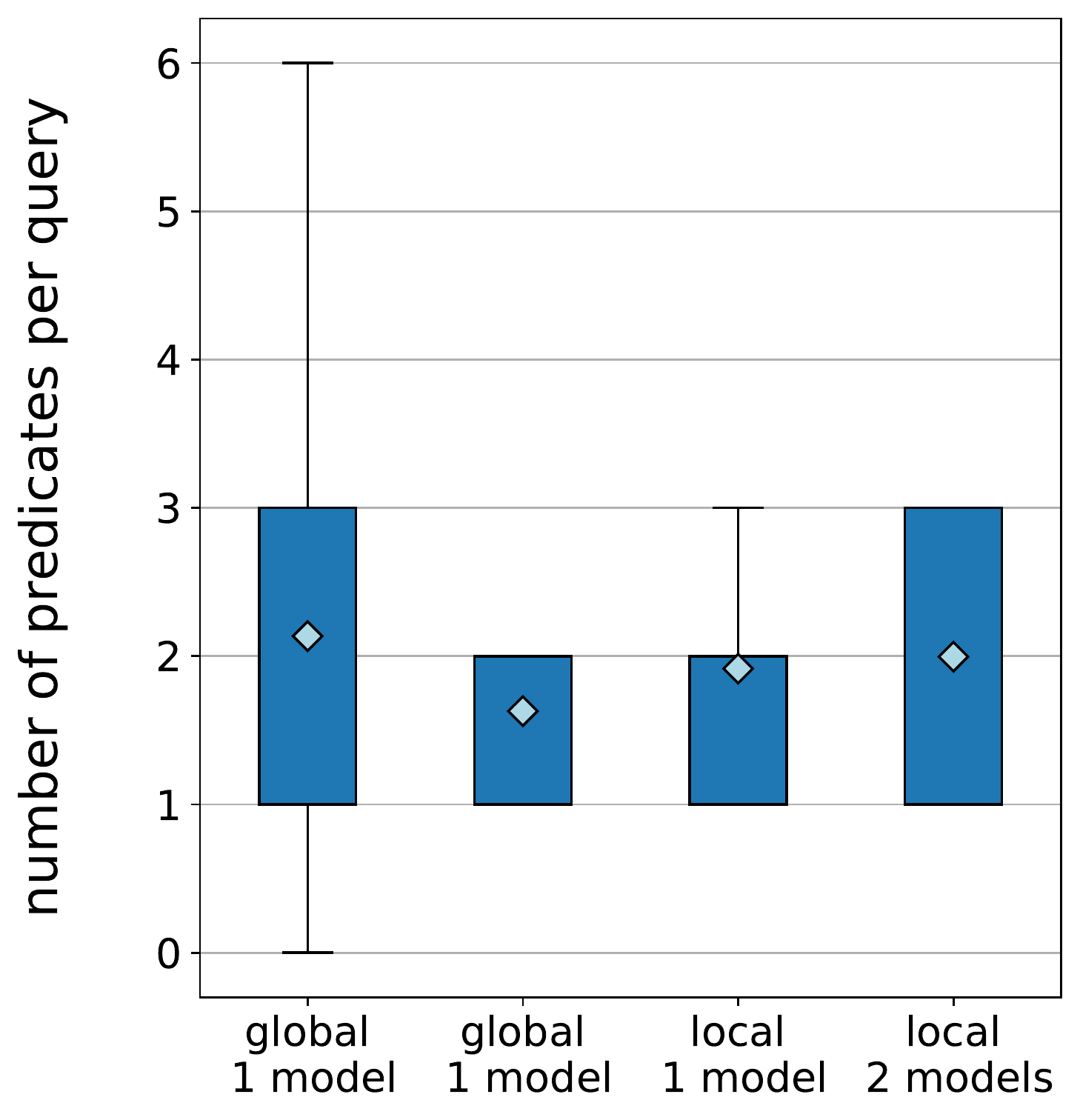}
        \caption{Number of predicates.}
        \label{fig:predicates}
    \end{subfigure}
    \caption{Analysis of ML-workload complexity for one global and 1 to 3 local models for the IMDB database.}
    \label{fig:complexity}
\end{figure*}

\subsection{Training Phase Workload Analysis}
\label{sec:analysis}
Fundamentally, the global as well as the local ML-based cardinality estimation approach use the same method to sample example queries from the context for the \emph{training phase}. 
This procedure is shown on the right side of Figure~\ref{fig:workload}, where a local model is trained on an example table consisting of three attributes \texttt{Brand}, \texttt{Color}, and \texttt{Year}. 
To train the local \emph{data-dependent} ML-model, a collection of count queries with different predicate combinations over the example table is generated. 
In our example, the predicates are specified over the three attributes using different operators $\neq,<,\leq,=,>,\geq$ and different predicate values.
This means, that all queries have the same structure but differ in their predicates to cover every aspect of the data properties in the underlying table. 
An example query workload to train a global model would look similar. 
However, the different contexts for global and local models have an impact on the number and the complexity of the example queries. 
In general, the query complexity is given by the combinations of joined tables and predicates in a query.
The larger the model context, the more complex the example queries.
Thus, global models have the highest workload complexity because they cover the whole schema.
Local models are trained with workloads with lesser complexity and fewer queries~\cite{woltmann2019local}.

\begin{figure}
    \centering
    \includegraphics[width=0.9\linewidth]{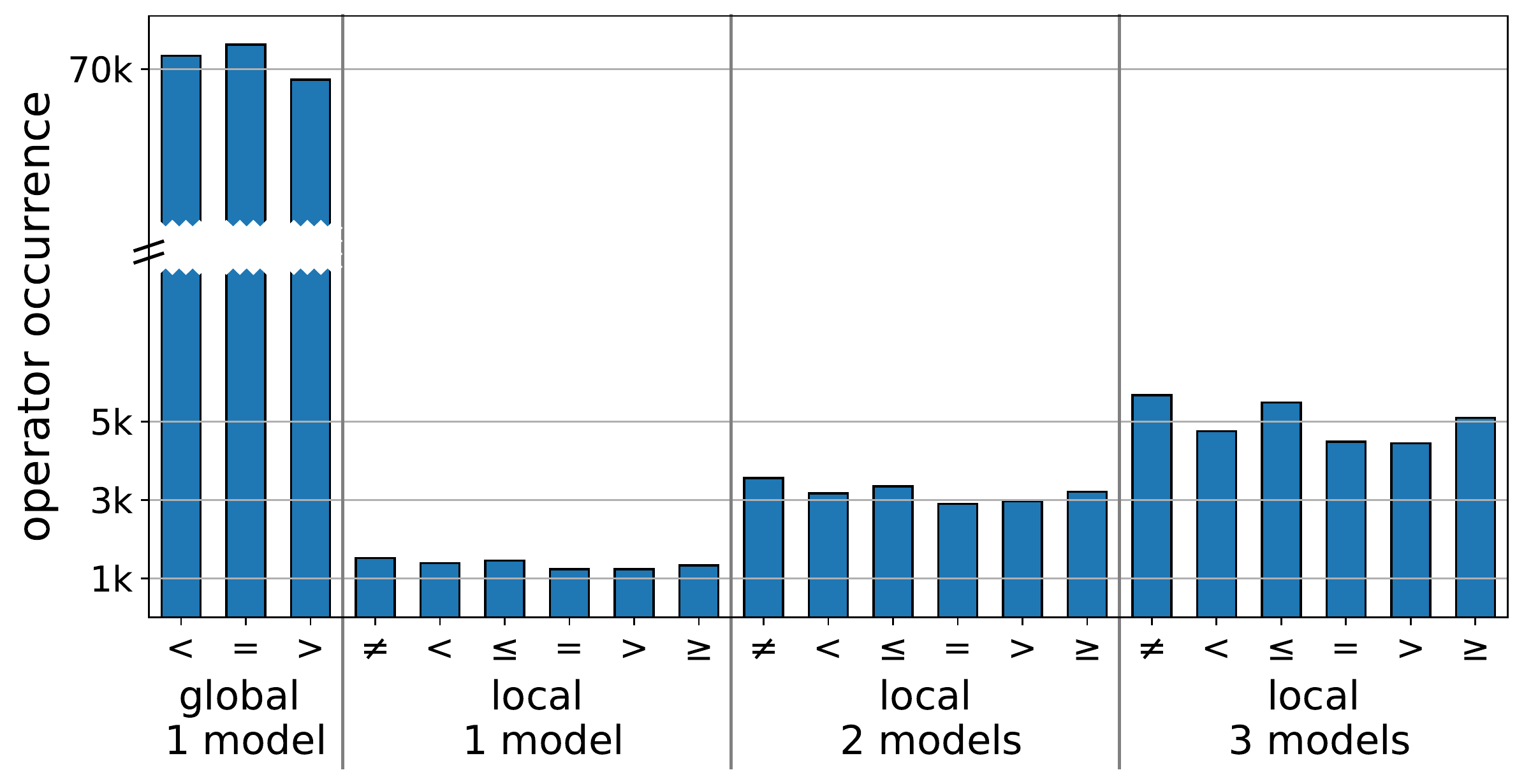}
    \caption{Predicate operator occurrences for IMDB.}
    \label{fig:operators}
\end{figure}

To better understand the query workload complexity for the training phase, we analyzed the workloads published by the authors of the global~\cite{global} and local approach~\cite{local}.
In both cases, the authors used the IMDB database~\cite{imdb} for their evaluation, because this database contains many correlations and thus, this is very challenging for cardinality estimation. 
The IMDB database contains more than 2.5 M movies produced over 133 years by more than 200,000 different companies with over 4 M actors. 
Our analysis results are summarized in Figure~\ref{fig:complexity} and \ref{fig:operators}. 

For one global model and an increasing number of local models, Figure \ref{fig:queries} shows the workload complexity in terms of numbers of example queries used per workload.
While the global model requires up to 100,000 example queries for the IMDB database~\cite{kipf2018learned}, the local model only requires 5,000 example query per local model~\cite{woltmann2019local} to determine a \emph{stable data-dependent} ML model.
In general, the number of example queries is much higher for a global model, but the number of example queries also increases with the number of local models.
Thus, several local models can be instantiated before their collective query count exceeds the number of queries for one global model.

Figure \ref{fig:tables} specifies the workload complexity in terms of data access through the number of accessed and joined tables per workload.
Similar to the previous figure, the global model has the highest complexity because it requires example queries over more tables to cover the whole schema at once.
Each local model covers only a limited part of the schema and therefore queries fewer tables per model.
Thereby, the complexity of accessed data for local models increases with the number of models.

Another important aspect to describe the workload complexity is the number of predicates per query as shown in Figure \ref{fig:predicates}.
Here, the global model workload has a much larger spread over the number of predicates.
The local models detail a more focused distribution with little variation.
Again, the local model workload does not require the amount of alternation in predicates of a global model workload because it covers less complexity of the schema.
In addition, the local model workload complexity still rises with more instantiated models.
In more detail, Figure \ref{fig:operators} gives an overview of the distribution of occurrences of all predicate operators in the workloads as a box plot with mean values.
The global model workload only uses the operators $<,=,>$, whereas the local model workloads use the full set of operators $\neq,<,\leq,=,>,\geq$.
As described by both authors, the predicate operators in each example query are sampled from a uniform distribution~\cite{kipf2018learned, woltmann2019local}.
The slight variation between the operators per workload is due to the fact that both approaches filter 0-tuple queries which do not occur uniformly.

\section{Training on Pre-Aggregated Data}
\label{sec:aggregation}

As discussed above, the global, as well as local ML-based approach for cardinality estimation, generates a lot of example queries with a count aggregate function during the \emph{training phase}.
Depending on the model context, there is a small variance in the number of accessed tables, but there is a high variance for predicates in terms of (i) number of predicates, (ii) used predicate operators, and (iii) predicates values in general. 
Additionally, the distribution of all these properties is uniform~\cite{kipf2018learned, woltmann2019local} which also adds to the variability in the specific query workload.
That means, that many queries work on the same data but look at different parts and computing count aggregation on every single part. 
Executing such workloads in a na\"ive way, i.e. executing each example query individually on large base data, is very expensive and generates a high load in the database system. 
The utilization of index structures for an optimized execution in database systems appears to be an ideal optimization technique at a first glance. 
However, their benefit is limited as we will show in our evaluation. 

To tackle that problem more systematically, our core idea is to \emph{pre-aggregate} the base data for different predicate combinations and to reuse this \emph{pre-aggregated} data by several example queries.
In general, aggregates compress the data by summarizing information and reducing redundancy.
This lessens the amount of data to be scanned by each example query because the aggregates are smaller than the original data.
The aggregate pre-calculates information with the result that the workload queries need to scan less data during execution.
Therefore, it is important that the construction of the aggregate must not take longer than the reduction of the workload execution time.



\begin{figure}
    \centering
    \includegraphics[trim=10 90 10 90, clip, width=0.9\columnwidth]{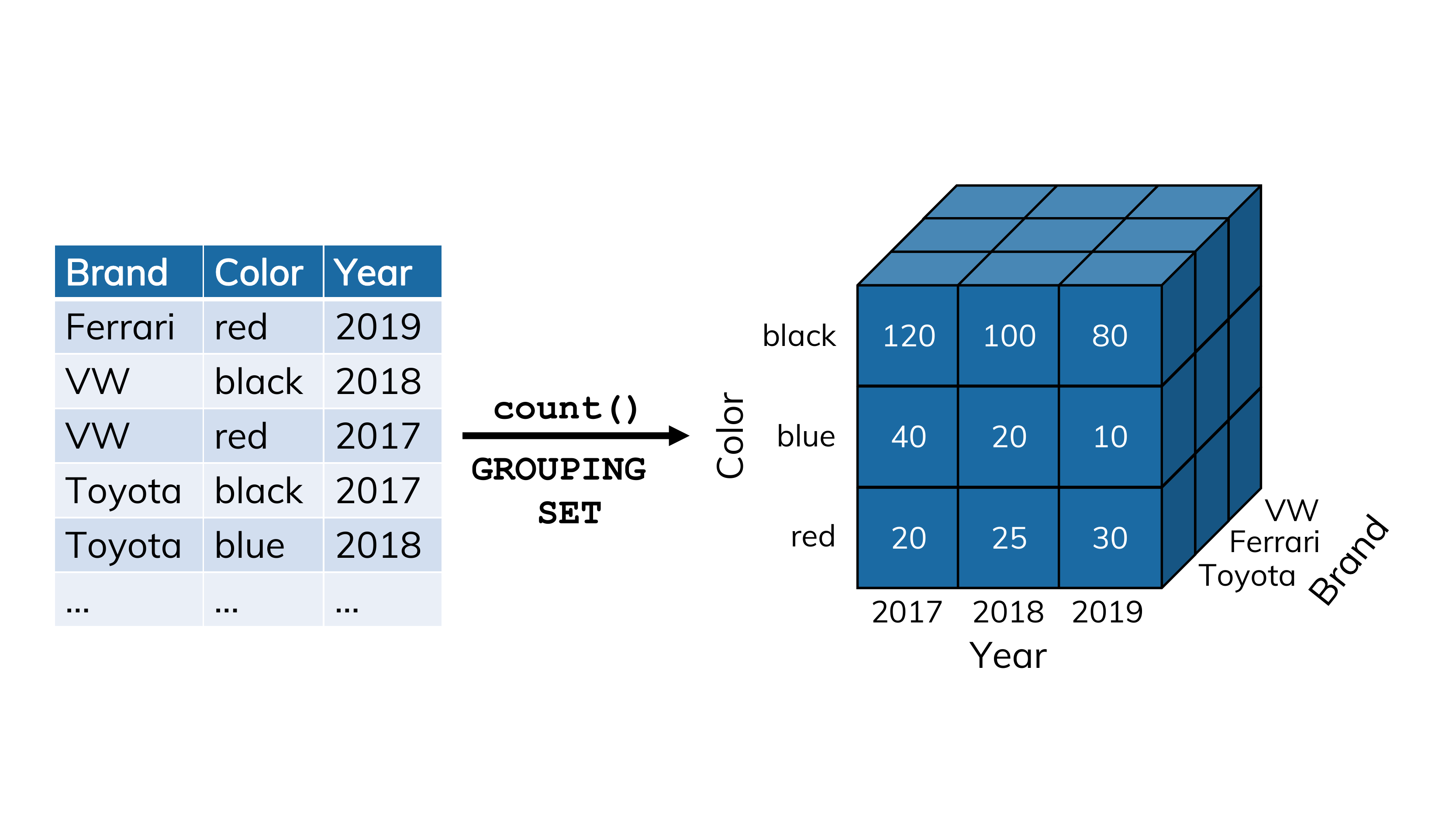}
    \caption{Aggregating information with grouping sets.}
    \label{fig:groupingsets}
    \vspace{-0.4cm}
\end{figure}

\subsection{Grouping Sets as Pre-aggregates}
It might sound expensive to aggregate all possible combinations of predicates.
However, DBMS already offer substantial supportive data structures for this kind of aggregation.
These data structures are able to group by all combinations of distinct values from a list of columns for a given aggregate function.
The basic idea of such grouping comes from \emph{Online Analytical Processing} (OLAP) workloads.
These aggregate-heavy workloads spawned the idea of pre-aggregating information in \emph{data cubes}~\cite{gray1996cube}.
These help to reduce the execution time of OLAP queries by collecting and compressing the large amount of data accessed into an aggregate.
The concept of data cubes is well-known from data warehouses for storing and computing aggregate information and almost all database systems are offering efficient support for data cube operations~\cite{DBLP:conf/vldb/AgarwalADGNRS96,gray1996cube,DBLP:conf/sigmod/HarinarayanRU96,shukla1996storage,DBLP:conf/sigmod/ZhaoDN97}.

Each attribute of a table or join generates a dimension in the data cube.
Moreover, the distinct attribute values are the dimension values.
The cells of a data cube are called facts and contain the aggregate for a particular combination of attribute values.
To instantiate the concept of a data cube in a DB, there are different \emph{cube operators}.
Usually, these are \texttt{CUBE}, \texttt{ROLLUP}, and \texttt{GROUPING SET}.
The \texttt{CUBE} operator instantiates aggregates for the power set of combinations of attribute values.
This includes combinations where not all attributes are used.
The \texttt{ROLLUP} operator builds the linear hierarchy of attribute combinations.
The hierarchy is determined by the order of the attributes.
The \texttt{GROUPING SET} operator only constructs combinations with all attributes.
This unique characteristic of a grouping set is the major advantage for our use case.
With the grouping set aggregation, we compress the original data and avoid the calculation of unnecessary attribute combinations.
Figure \ref{fig:groupingsets} details an example of a grouping set for a count aggregate over discrete data about cars.
The example data has a multidimensional structure after aggregating where each dimension is a property of a car.
The cells of the grouping set are filled with the aggregate value, i.e. the count of cars with a particular set of properties.

\begin{figure}
    \centering
    \includegraphics[width=0.95\columnwidth]{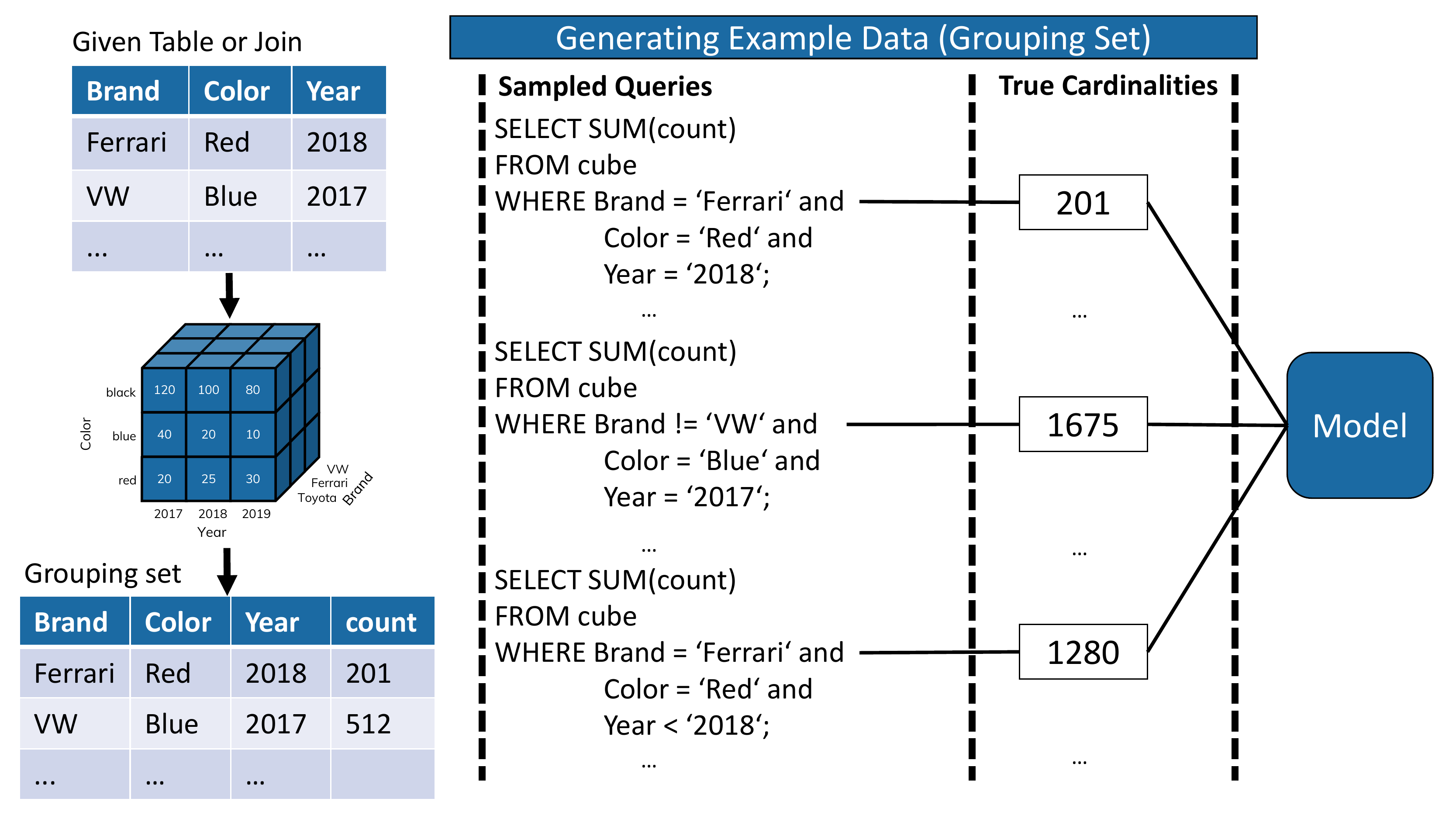}
    \caption{Illustration of database-supported training phase based on pre-aggregated data using grouping-sets.}
    \label{fig:optworkload}
    \vspace{-0.4cm}
\end{figure}

\begin{figure*}
    \centering
    \begin{subfigure}[b]{0.9\textwidth}
    \centering
        \includegraphics[width=\textwidth]{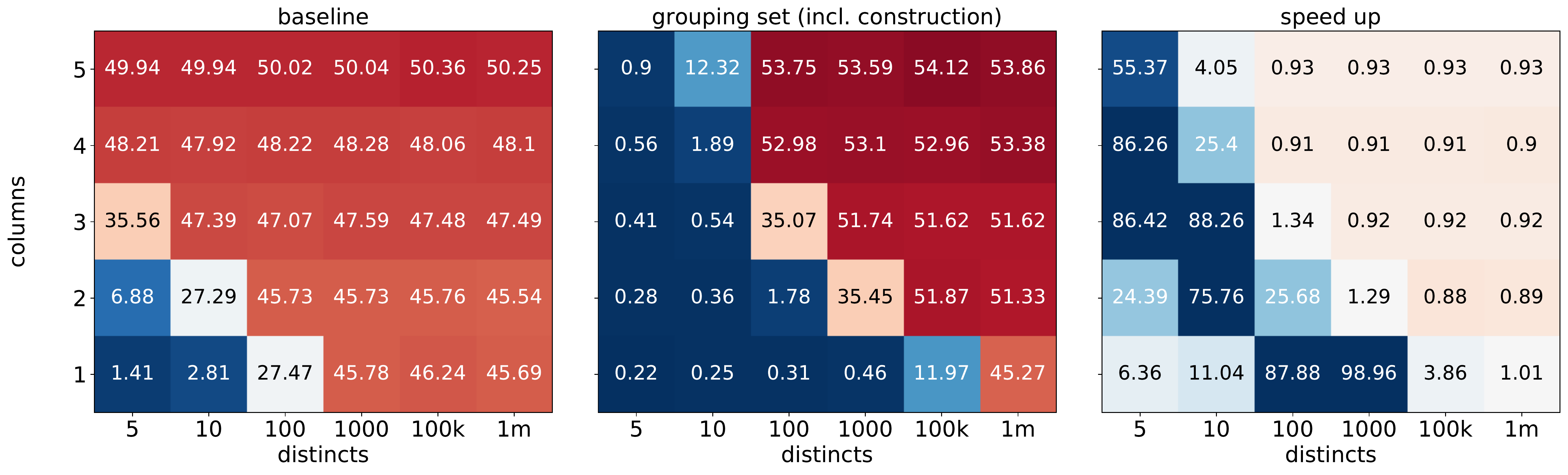}
        \caption{The speed up for different combinations of columns and distinct values with tuples fixed to 1m.}
        \label{fig:tuples}
    \end{subfigure}
    \begin{subfigure}[b]{0.9\textwidth}
    \centering
        \includegraphics[width=\textwidth]{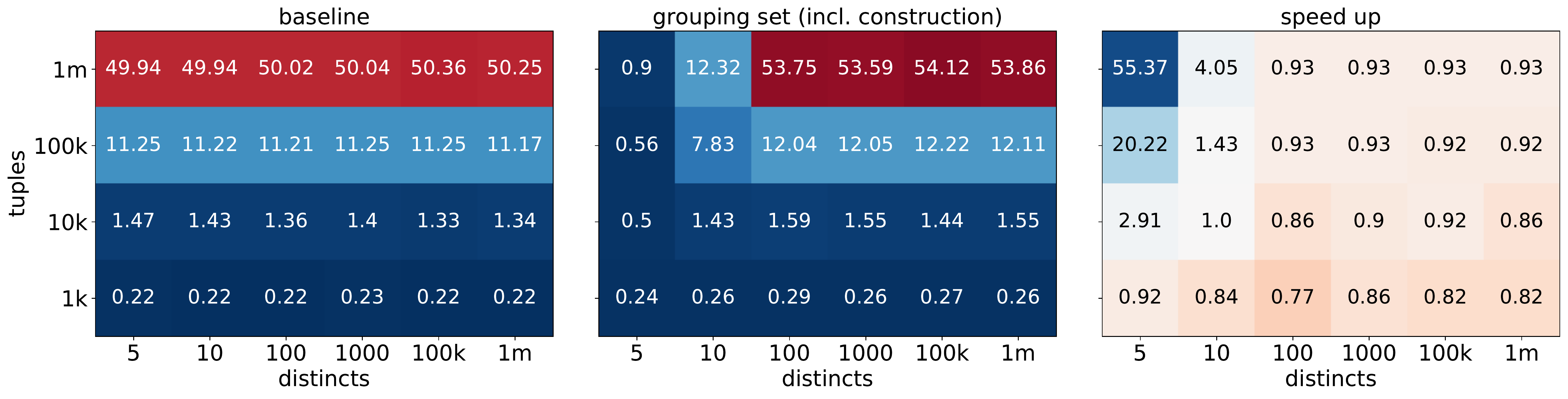}
        \caption{The speed up for different combinations of tuples and distinct values with columns fixed to five.}
        \label{fig:columns}
    \end{subfigure}
    \begin{subfigure}[b]{0.9\textwidth}
    \centering
        \includegraphics[width=\textwidth]{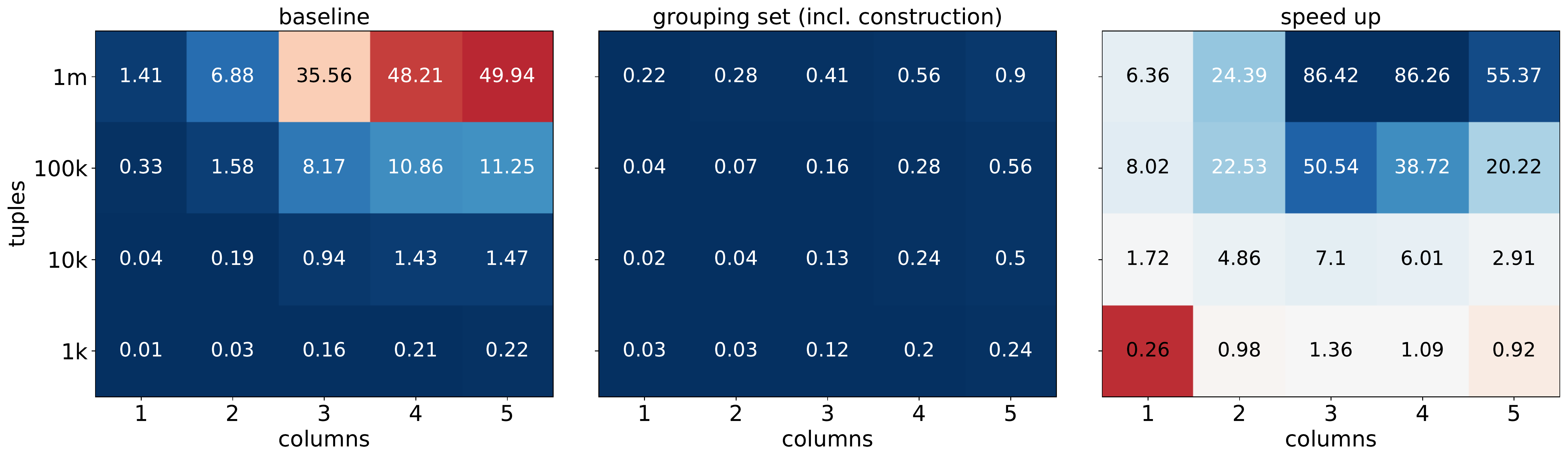}
        \caption{The speed up for different combinations of tuples and columns with distinct values fixed to five.}
        \label{fig:distinct}
    \end{subfigure}
    \caption{Execution times and speed ups on synthetic data.}
    \label{fig:synthetic}
    \vspace{-0.4cm}
\end{figure*}

Given the grouping set data structure, we adapt the general generation of example queries from Figure \ref{fig:workload}.
By introducing the data cube, we add an intermediate step before executing the workload.
This step constructs a data cube, i.e. grouping set, and rewrites the workload to fit the grouping set.
Figure \ref{fig:optworkload} details the construction and the rewrite of queries for the car data.
On the left side, the construction builds a table matching the theoretical multidimensional character of the grouping set.
Due to this new table layout, the rewrite must include a different aggregate function as shown in Figure \ref{fig:optworkload}.
For a count aggregate, the corresponding function is a sum over the pre-aggregated count.
Last, the rewritten workload is executed and retrieves the \texttt{output-cardinalities}.
After the adapted sampling of example queries, the queries and cardinalities are fed to the ML model in the same way as in the original process.
This is a great advantage of our approach because it does not interfere with other parts of the training process.
Therefore, it is independent of the type of ML model and can be applied to a multitude of supervised learning problems. 

Even though our approach is independent of the ML model, it is not independent of the data.
In general, grouping sets are only beneficial if the aggregate is smaller than the original data.
Otherwise, the queries would not benefit from the pre-aggregation.
A negative example would be an aggregate over several key columns.
Here, the number of distinct values per column equals the number of tuples in the table.
If such a grouping set is instantiated, each of its dimensions has a length equal to the number of tuples.
This grouping set is larger than the original data.
Thus, a query on this aggregate runs longer because it actually has to scan more data.
With all that in mind, we need to quantify the benefit of a grouping set for our use cases.

\subsection{Benefit Criterion}
To find a useful criterion when to instantiate grouping sets, we evaluate the usefulness of these sets on synthetic data.
Again, we need to find a way to express the compression of information in a grouping set.
From the synthetic data, we will derive a general rule for the theoretical improvement of a grouping set on a given table or join.
Our experiment comprises six steps per iteration.
The first step generates a synthetic table given three properties.
These properties are: the number of tuples in a table or join $N$, the number of columns $C$, and the number of distinct values per column $distinct$.
We vary the properties in the ranges:
\begin{align}
    N &\in \{1\,000, 10\,000, 100\,000, 1\,000\,000\}\\
    C &\in \{1, 2, 3, 4, 5\} \\
    distinct &\in \{5, 10, 100, 1\,000, 100\,000, 1\,000\,000\} 
\end{align}
Exactly one property is changed in a single iteration.
This leads to $4 \cdot 5 \cdot 6 = 120$ different tables or iterations.
The values in a column are uniformly sampled from the range of distinct values.
With an increasing number of distinct values per column, we simulate columns of type floating point which have a large number of different values.
The other way around, columns with few distinct values resemble categorical data.
In the second step, we sample 1,000 count aggregate queries as an example workload over all possible combinations of columns (predicates), operators, and values in this iteration.
In the third step, we execute these queries against the table and measure their execution time.
This is equivalent to the standard procedure to sample example cardinality queries for an ML model.
The fourth step constructs the grouping sets over the whole synthetic table and measures its construction time.
Next, in step five, we rewrite the queries in a way that they can be executed against the grouping set.
We measure their execution time on this grouping set.
In the final step, we divide the execution time of the workload on the grouping set by the run time of the workload on the table.
We call this quotient the \emph{speed up factor}.
It ranges from close to zero for a negative speed up to infinity for a positive speed up.
A speed up factor of one would mean no change.
All time measurements are done three times and averaged.
We use PostgreSQL 10 for the necessary data management.

Figure \ref{fig:synthetic} shows the results of all 120 iterations.
Blue values mean either better execution times or higher speed up, whereas red means longer execution times or lower speed up.
White indicates a speed up factor of one.
The first column shows the execution time of the workload against the table.
The next column shows the execution time of the workload against the grouping set including the construction time of the grouping set.
The last column is the quotient of the second and first column.
This is the achieved speed up by using a grouping set.
In each row, only two properties are changed while the third property is kept fixed.
The first row keeps the number of tuples, the second row the number of columns, and the last row the number of distinct values fixed.
From this figure, we can derive three conclusions.

First, we notice that few distinct values in a few columns are beneficial for the aggregation.
Next, the more tuples $N$ in a table, the more distinct values per column can be there for the speed up to be sustained.
As the last conclusion, this also applies to the number of columns.
All in all, the larger the original table the more distinct values and columns still lead to a speed up.
Our experiments show that a grouping set is only beneficial if its size is smaller than the original table.
Only then the aggregate compresses information and causes less data to be scanned by the queries.
Given our evaluation, this happens if the product of the distinct values of all columns is smaller than the table size.
We can model this as an equation to be used as a criterion for instantiating beneficial grouping sets.
\begin{equation}
    \text{scaling factor} = \frac{1}{N}\prod\limits_{c=1}^{C} \left|\text{distinct\_values} (\text{column}_c)\right|
    \label{eq:scaling}
\end{equation}
If this \emph{scaling factor} is smaller than one, we call a grouping set beneficial.
The scaling factor is also a measure of data compression.
Therefore, it shows how much faster the scan over the aggregated data can be.


\begin{algorithm}[t]
    \DontPrintSemicolon
    \SetKwInOut{Input}{Input}\SetKwInOut{Output}{Output}
    \SetKw{KwTo}{in}\SetKw{Of}{of}\SetKw{Where}{where}
    \SetKwFunction{Construct}{construct}\SetKwFunction{Max}{max}
    \SetKwFunction{Add}{add}
    \Input{$\,$ workload $wl$, database}
    \Output{$\,$ grouping sets}
    \BlankLine
    grouping sets: tables $\rightarrow$ attributes
    \BlankLine
    \For(\tcp*[f]{Analyze}){query q \KwTo wl}{
        tables, attributes \Of q\tcp*[f]{step 1}\\
        
        grouping sets[tables] =  \\
        $\phantom{grouping}$grouping sets[tables] $\cup$ attributes\\
    }
    \BlankLine
    \For{grouping set gs \KwTo grouping sets}{
        N = \Max(|tuples| \Of all tables \Of gs)\tcp*[f]{step 2}\\
        \For{attribute p \KwTo attributes \Of gs}{
            dv = dv $\cup$ |distinct values \Of p|\\
        }
        
        scaling factor = $\frac{\prod\text{dv}}{N}$\tcp*[f]{step 3}\\
        \uIf{scaling factor $\ge 1$}{
            grouping sets[tables] = attributes \Where $\frac{\prod\text{dv}}{N} < 1$
        }
        
        \Construct grouping set
    }
   \caption{Analyze component.}
    \label{alg:analyze}
\end{algorithm}

\section{Implementation}
\label{sec:training}

In this section, we describe the implementation of our \emph{aggregate-enabled training phase} for ML-based cardinality estimation in DBMS in detail. 
In our implementation, we assume that a regular DBMS with an SQL interface provides the base data and the ML models are trained outside the DBMS, e.g., in Python with keras\footnote{\url{https://keras.io}}. 
Based on this setting, we added a new layer implemented in Python between these systems to realize our \emph{aggregate-enabled training phase} in a very flexible way.
Thus, the input of this layer is an ML workload that is necessary for training the ML model. 
Then, the main tasks of this layer are:
\begin{compactenum}
\item discover as well as create as many beneficial grouping sets in DBMS as possible for the given ML workload and
\item rewrite as well as execute the workload queries according to the grouping sets and base data. 
\end{compactenum}
The output of this layer is an annotated ML workload with the determined cardinalities on which the ML model is trained afterward.
To achieve that, our layer consists of two components. 
The first component is the \texttt{Analyzer} which is responsible for the construction of beneficial grouping sets. 
The second component is the \texttt{Rewrite} rewriting and executing the queries of the ML workload against the constructed grouping sets. 
In the following, we introduce both components in more detail.

\subsection{Analyzer Component}
Algorithm~\ref{alg:analyze} gives a more detailed overview over the \texttt{Analyzer Component}. 
Given an ML workload and a database, our \texttt{Analyzer} consists of three steps to find and build all beneficial grouping sets. 
In \textbf{step one}, the \texttt{Analyzer} scans all queries in the ML workload and collects all joins or tables and their respective predicates in use.
This generates all possible grouping sets as a mapping from tables building the grouping set to the predicates on those tables.
In Algorithm \ref{alg:analyze}, this is covered in lines one to 6.
Then, the \textbf{second step} collects the number of distinct values per predicate attribute and the maximum number of tuples of all tables in the grouping set from the metadata (statistics) of the database.
This step is shown in lines 8 to 11.

In the \textbf{third and final step}, our defined benefit criterion (Equation \eqref{eq:scaling}) is used to calculate the scaling factor and therefore the benefit of each grouping set.
If the scaling factor is smaller than one, the Analyzer constructs the grouping set with all collected predicates.
If the scaling factor is larger than or equal to one, the grouping set is constructed with the maximum number of predicates where the scaling factor still is smaller than one.
This may disregard queries not to be executed against the grouping set if they have more predicates than the grouping set.
On the other hand, queries on the table or join with the predicates in the grouping set can still benefit from it.
Moreover, all queries to be executed against a grouping set are marked for rewriting.
This final step is detailed in lines 12 to 15. 

\begin{figure*}
    \begin{subfigure}[t]{0.33\textwidth}
        \centering
        \includegraphics[height=4.5cm]{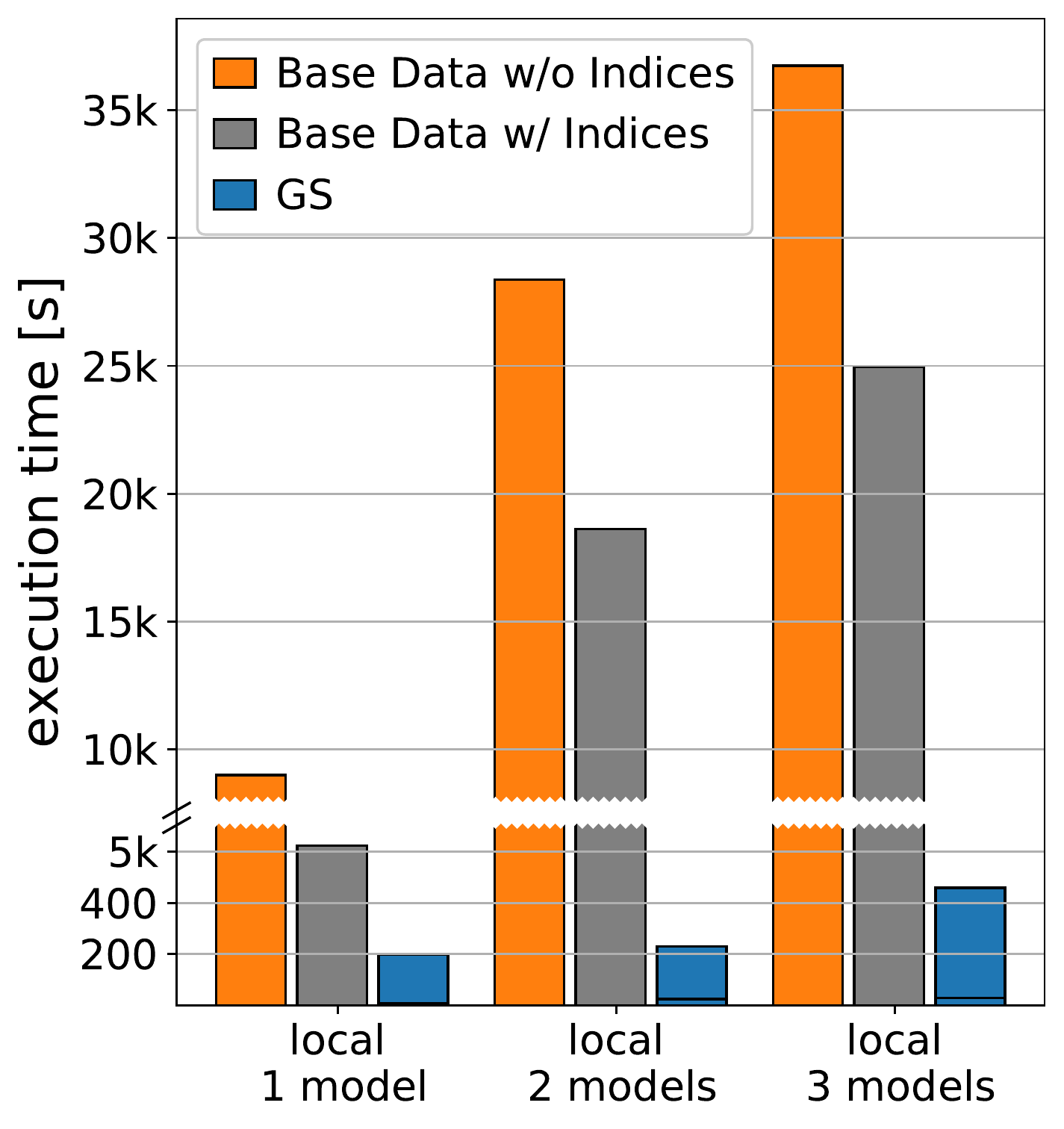}
        \caption{Execution times.}
        \label{fig:execlocal}
    \end{subfigure}
    \begin{subfigure}[t]{0.33\textwidth}
        \centering
        \includegraphics[height=4.5cm]{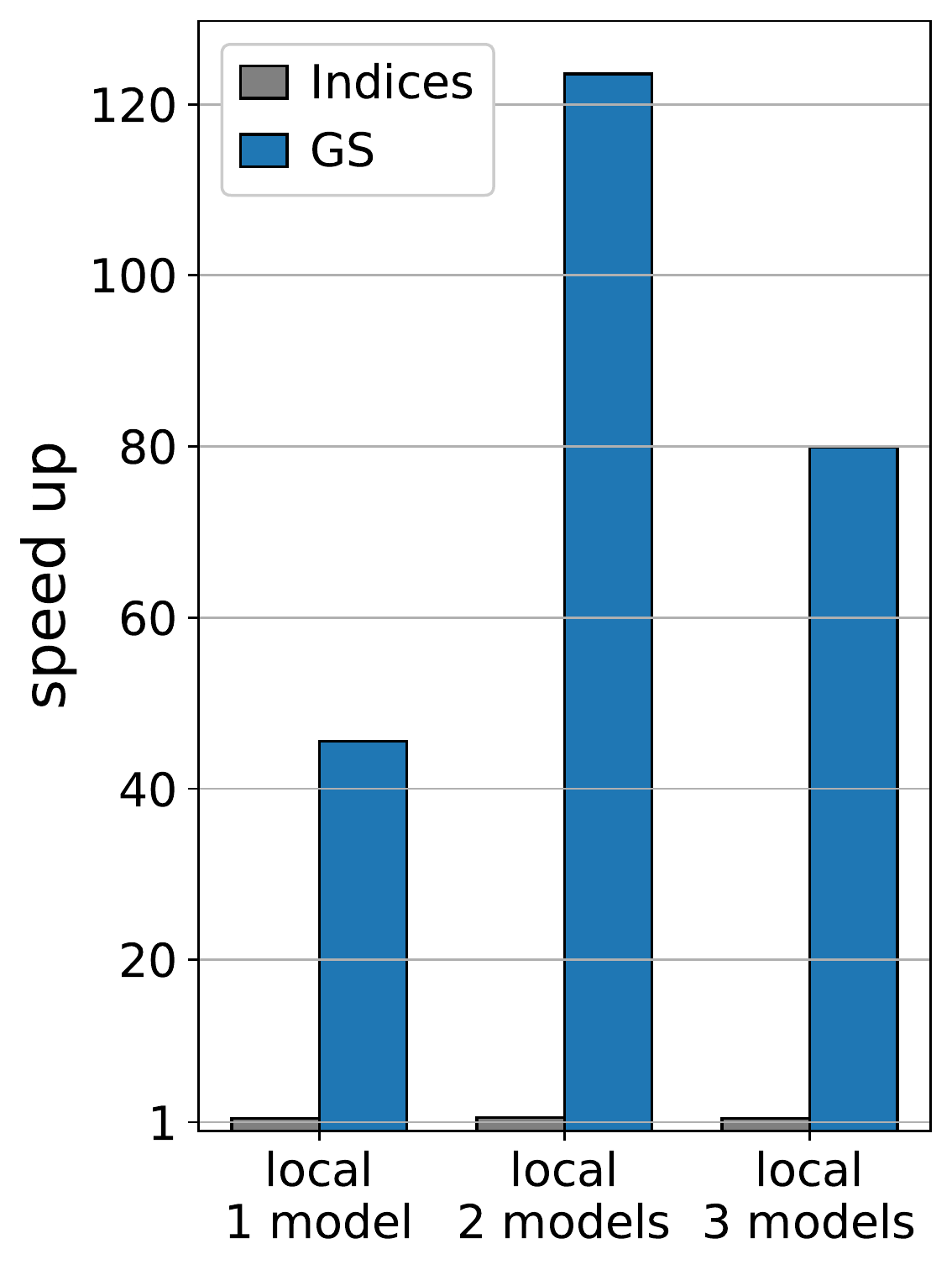}
        \caption{Speed ups.}
        \label{fig:speedlocal}
    \end{subfigure}
    \begin{subfigure}[t]{0.33\textwidth}
        \centering
        \includegraphics[height=4.5cm]{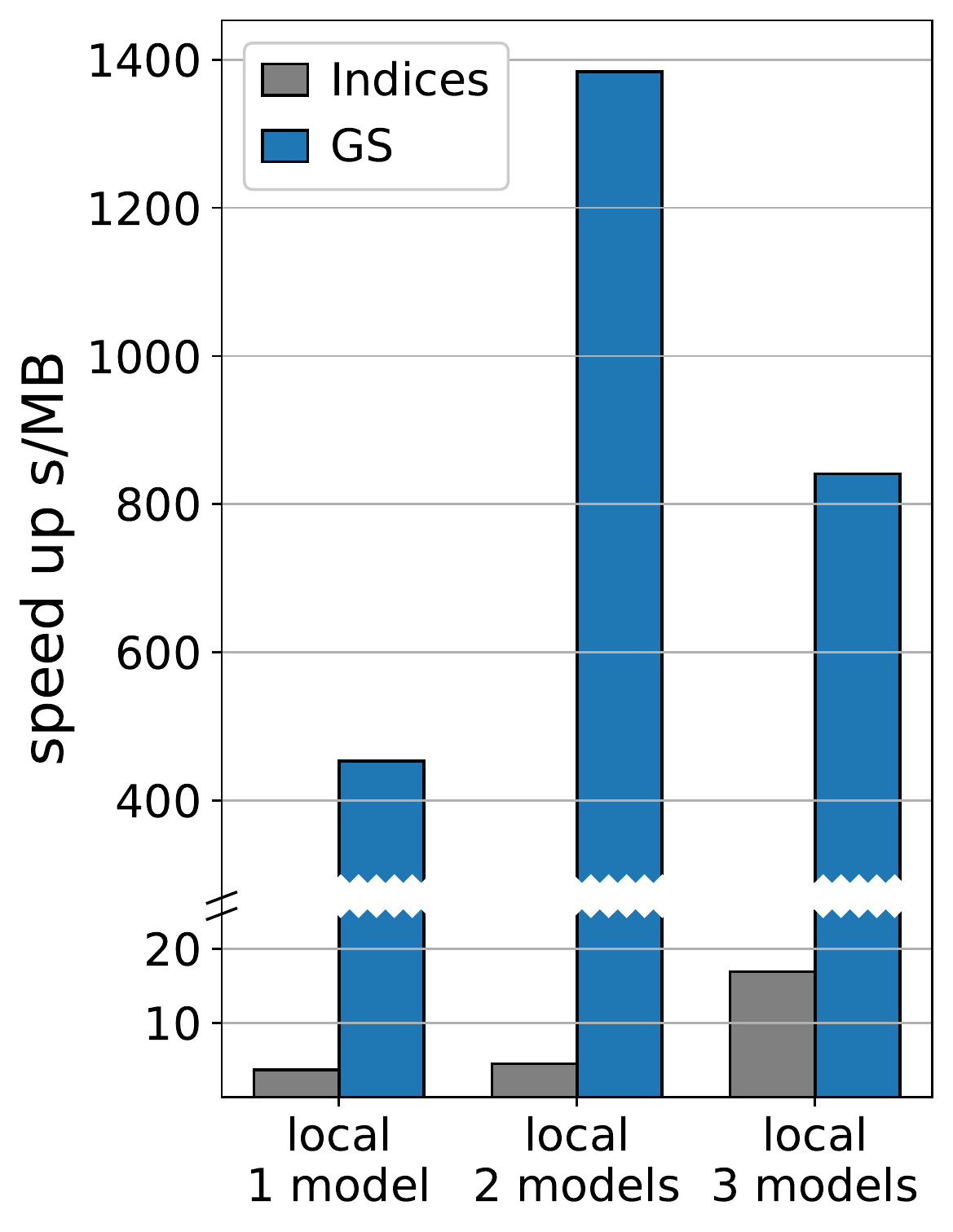}
        \caption{Speed up per MB.}
         \label{fig:memlocal}
    \end{subfigure}
    \vspace{-0.3cm}
    \caption{Local model evaluation results based on our \emph{aggregate-enabled training phase} (GS: Grouping Sets).}
    \label{fig:local}
    \vspace{-0.4cm}
\end{figure*}

\begin{algorithm}[t]
    \DontPrintSemicolon
    \SetKwInOut{Input}{Input}\SetKwInOut{Output}{Output}
    \SetKw{KwTo}{in}\SetKw{Of}{of}\SetKwFunction{Rewrite}{rewrite}
    \SetKwFunction{Add}{add}
    \Input{$\,$ workload $wl$, grouping sets}
    \Output{$\,$ rewritten workload $wl$'}
    \BlankLine
    \For(\tcp*[f]{Rewrite}){query q \KwTo wl}{ 
        tables, attributes \Of q\\
        
        \uIf{attributes $=$ grouping sets[tables]}{
            \Rewrite q to match grouping set
        }
        \Add q to $wl$'
    }
   \caption{Rewrite component.}
    \label{alg:rewrite}
\end{algorithm}

\subsection{Rewrite Component}
With all beneficial grouping sets instantiated by the \texttt{Analyzer Component}, it is necessary to modify the ML workload queries to be able to use the pre-aggregates.
For this, all queries which can be run against any grouping set will be rewritten in the \texttt{Rewrite} component.
The \texttt{Rewrite} component receives information about each query from the \texttt{Analyzer} and rewrites queries in a way that they can be executed against the grouping sets.
All queries where the \texttt{Analyzer} does not recognize a grouping set are kept as they are and will be executed over the base data.
The \texttt{Rewrite} component is described in Algorithm \ref{alg:rewrite}.

When all queries have been processed, the optimized workload is executed as a whole on the database as the last step. 
If a query has been rewritten, it will be executed against the grouping set, otherwise, it will be executed against the original data.
Finally, the determined results (i.e. cardinalities) are forwarded to the ML system to train the ML model.

\section{Evaluation}
\label{sec:eval}

\begin{table*}
    \centering
    \begin{tabular}{@{}lrrrrrr@{}}
        \toprule
        model & Base Data w/o Index & Base Data w/ Index & Construction GS & Execution GS & Total & Coverage GS\\
        \hline
        local 1 & $2h\,29m\,57s$ & $1h\,44m\,01s$ & $6.17s$ & $191.34s$ & $197.51s$ & $5\,000/5\,000 = 100\%$\\
        local 2 & $7h\,52m\,52s$ & $5h\,10m\,27s$ & $23.70s$ & $205.91s$ & $229.61s$ & $10\,000/10\,000 = 100\%$\\
        local 3 & $10h\,12m\,24s$ & $6h\,55m\,55s$ & $29.10s$ & $430.36s$ & $459.46s$ & $15\,000/15\,000 = 100\%$\\
        \hline
        global full & -- & $4d\,14h\,11m$ & $2h\,22m\,20s$ & $20d\,20h\,02m$ & $20d\,22h\,24m$ & $100\,000/100\,000 = 100\%$\\
        global opt & -- & $4d\,14h\,11m$ & $34m\,29s$ & $2d\,11h\,03m$ & $2d\,11h\,38m$ & $54\,722/100\,000 =\,\; 55\%$\\

        \bottomrule
    \end{tabular}
    \caption{Execution times ML workloads (gs: grouping sets).}
    \label{tab:runtimes}
    \vspace{-0.4cm}
\end{table*}

To show the benefit of our novel \emph{aggregate-enabled training phase}, we conducted an exhaustive experimental study with both presented types of ML models for cardinality estimation (cf. Section~\ref{sec:models}). 
Thus, we start this section by explaining the experimental settings followed by a description of selective results for the local as well as global ML model approaches. 
Afterward, we summarize the main experimental findings. 

\subsection{Experimental Setting}

For our experiments, we used the original workloads for the local and global ML model approaches~\cite{global, local} on the IMDB data set~\cite{imdb}.
The IMDB contains a snowflake database schema with several millions of tuples in both the fact and the dimension tables.
As already presented in Section~\ref{sec:analysis}, the global model workload contains 100,000 queries. 
The local model workload consists of three consecutive workloads where each workload has 5,000 queries more than the previous one.
This corresponds to an increase of one local model per workload. 
That means, we have four workloads for our experiments: one for a global model and three workloads for an increasing number of local models.
Moreover, all experiments are conducted on an AMD A10-7870K system with 32GB main-memory with PostgreSQL 10 as the underlying database system.

In our experiments, we measured the workload execution times, whereby we distinguish three different execution modes:
\begin{compactenum}
\item[\textbf{Base Data w/o Indexes:}] ML workload is executed on the IMDB base data without any index on the base data. 
\item[\textbf{Base Data w/ Indexes:}] ML workload is executed on the IMDB base data with indices on all columns in use.
\item[\textbf{Grouping Set (GS):}] ML workload is executed on \emph{pre-aggregated data} as determined by our approach.
\end{compactenum}
The first two execution modes are our baselines because both a currently used in the presented ML model approaches for cardinality estimation~\cite{kipf2018learned, woltmann2019local, liu2015nn}.   
Moreover, the workloads are executed for all three different execution modes using the Python-PostgreSQL bridge. 


\subsection{Experimental Results}

We present the results for each ML model approach individually. 

\subsubsection*{\textbf{Local Model Workloads}}
Figure~\ref{fig:local} shows the results for the local model workloads, whereas we used three workloads. 
The first workload contains the necessary queries to build one local ML model to estimate the cardinalities for the join \sql{title}$\bowtie$\sql{movie\_keyword}.
The second workload adds 5,000 queries to the first workload to construct a second ML model for the join  \sql{title}$\bowtie$\sql{movie\_info}. 
The third workload adds another ML model for an additional join \sql{title}$\bowtie$\sql{movie\_companies}.
That means we increase the number of local ML models per workload by one. 

Figure \ref{fig:execlocal} details the execution times for all three local model workloads for all investigated execution modes.
In each of the three groups, the left bar shows the complete workload execution time on the IMDB base data w/o index, the middle bar on the IMDB base w/ index, and the right bar the execution time with our grouping set approach. 
As we can see, indexes on the base data are already helping to reduce the workload execution times compared to execution on the base data w/o indexes, but the speedup is very marginal as shown in Figure~\ref{fig:speedlocal}. 
In contrast to that, our grouping set approach has the lowest execution times in all cases and the achieved speed ups compared to execution on the base data w/o indexes are in the range between 45 and 125 as depicted in Figure~\ref{fig:speedlocal}.
From that, we can conclude that our \emph{aggregation-enabled training phase} is much more efficient than state-of-the-art approaches. 

For each considered join, our \emph{aggregation-enabled} approach creates a specific grouping set containing all columns from the corresponding workload queries.
According to our definition, the scaling factors were: $0.0169$, $0.00257$, and $0.0524$ for the respective joins.
Thus, the instantiation of grouping set is beneficial. 
As we can see, the scaling factors are very low. 
So, all grouping sets achieve a very good compression rate and the rewritten workload queries on the grouping sets have to read much less data compared to the execution on base data. 
Moreover, all workload queries can be rewritten, so that the coverage is 100\% and every query benefits from this optimization. 
Nevertheless, the scaling factors differ, which explains the different speed ups. 

However, the construction of the grouping sets can be considered a drawback. 
Though, as illustrated in Table~\ref{tab:runtimes}, the construction times for the grouping sets are negligible because the reduction in execution time is significantly higher. 
From a storage perspective, index structures and grouping sets need some extra storage space, where the storage overhead for grouping sets is larger than for indices. 
But, as illustrated in Figure~\ref{fig:memlocal}, the speed up per additional MB for grouping sets is much larger than for indices. 
All in all, we gain a much larger speed up making grouping sets the more efficient approach.

\begin{figure}
    \begin{subfigure}[t]{0.23\textwidth}
        \includegraphics[height=5cm]{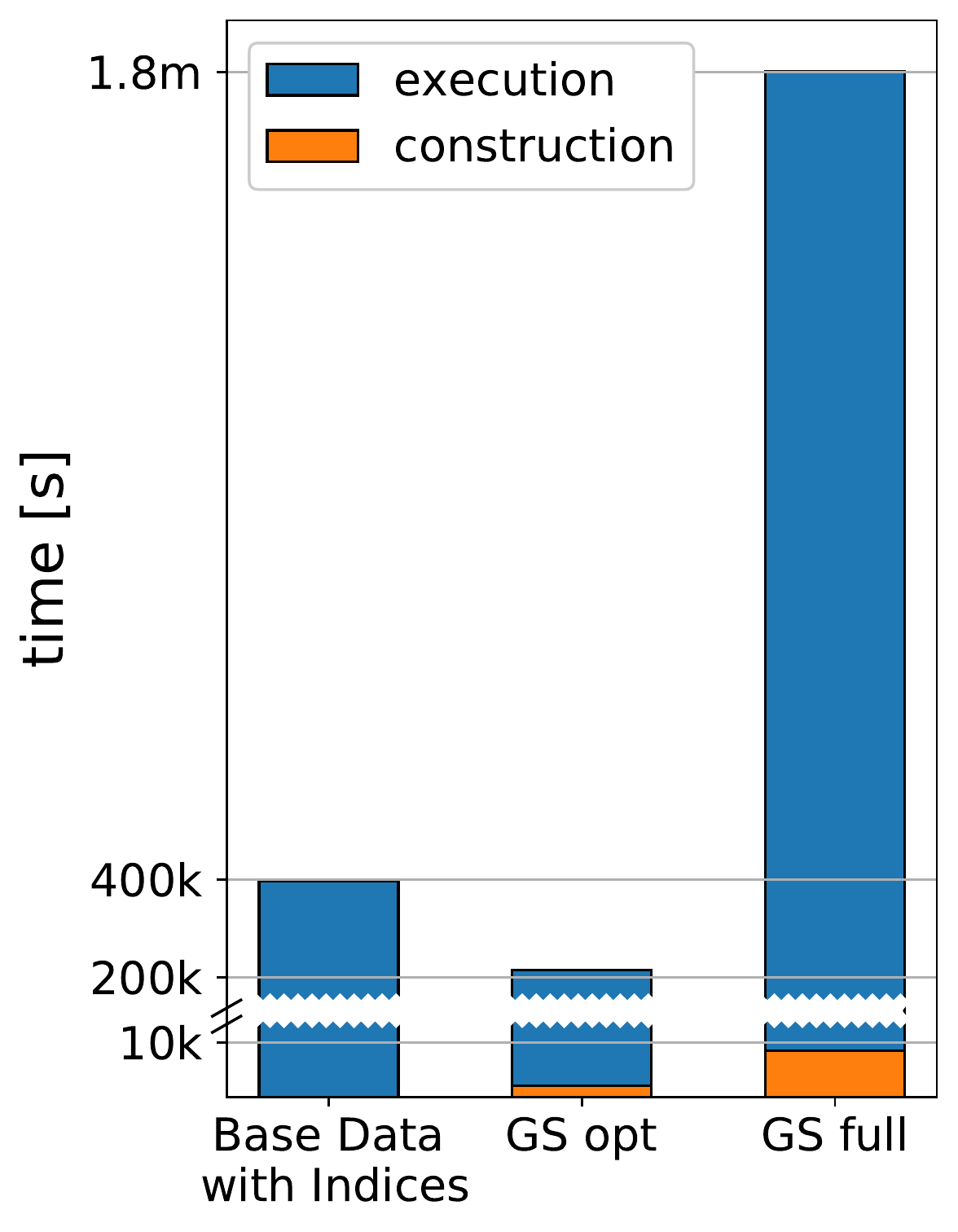}
        \caption{Execution times.}
        \label{fig:execglobal}
    \end{subfigure}
    \hfill
    \begin{subfigure}[t]{0.23\textwidth}
        \raisebox{6pt}{\includegraphics[height=4.8cm]{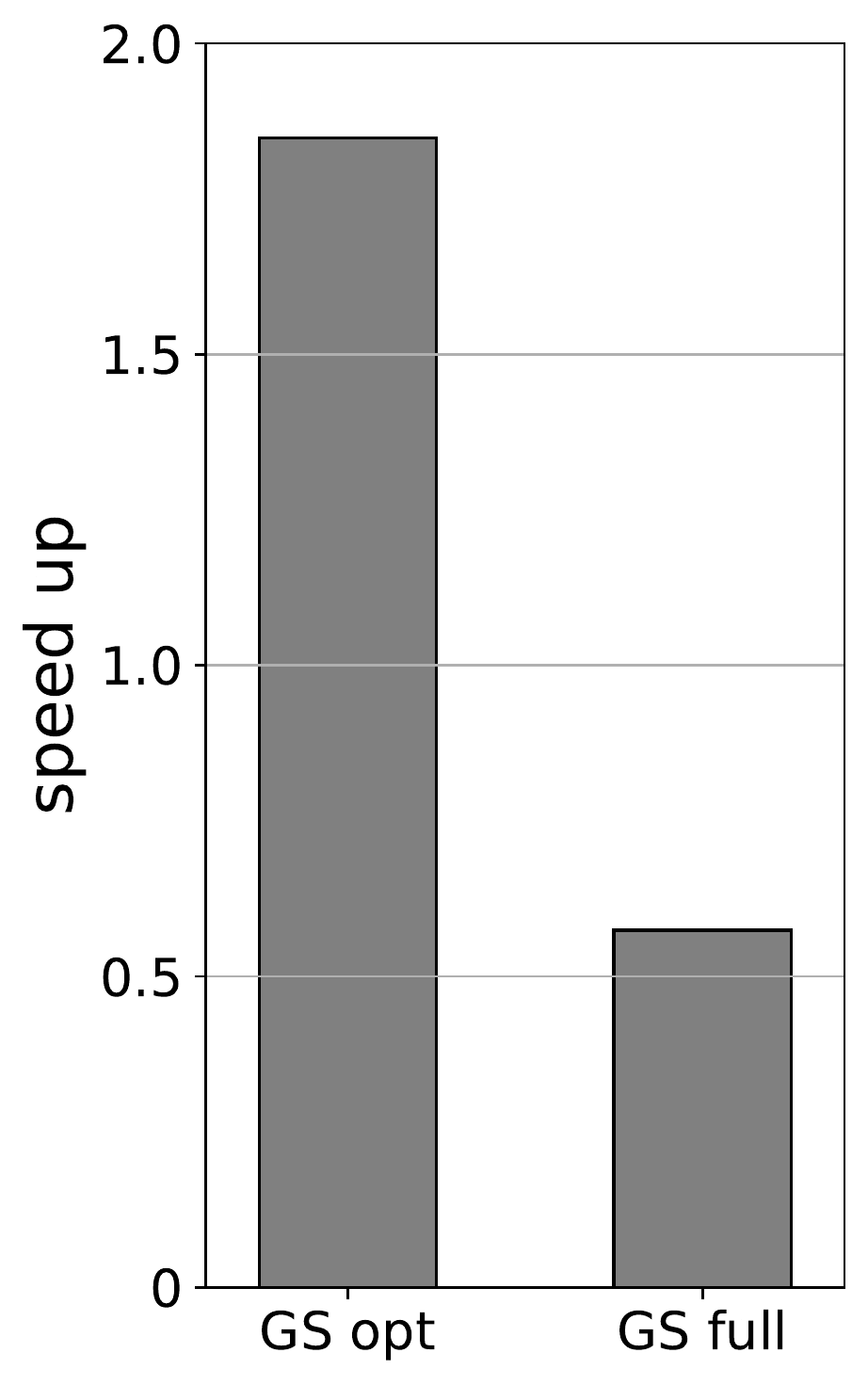}}
        \caption{Speed ups.}
        \label{fig:speedglobal}
    \end{subfigure}
        \vspace{-0.3cm}
    \caption{Global model evaluation results.}
    \label{fig:global}
    \vspace{-0.4cm}
\end{figure}

\subsubsection*{\textbf{Global Model Workload}}
Figure~\ref{fig:global} shows the evaluation results in terms of execution and construction times for the investigated global model workload. 
As shown in the previous experiment, the utilization of indices is always beneficial. 
Thus, we only compare the execution on base data w/ indices and the execution on the aggregated data in this evaluation. 

In general, there are 21 grouping sets possible for the global workload.
However, some of these grouping sets have a scaling factor larger than one.
Therefore, our \texttt{Analyzer} component, disregards the attributes of some grouping sets until the scaling factor is smaller than one (cf. Section~\ref{sec:training}). 
As a consequence, only 55\% of the global workload queries can be rewritten to this optimal set of grouping sets. 
The remaining 45\% have to be executed against the original data.
Nevertheless, this strategy called \emph{GS opt} in Figure~\ref{fig:execglobal} reduces the workload execution time of the global workload.
The speed up compared to the execution on the base data w/indices is almost 2 (Figure~\ref{fig:speedglobal}). 

To show the benefit of our grouping set selection strategy, we also constructed all grouping sets with all attributes (\emph{GS full}). 
There, we are able to rewrite all global workload queries to be executed on these aggregated data. 
As shown in Figure~\ref{fig:execglobal}, the overall workload execution time is larger than the execution on base data w/ indices. 
Therefore, grouping sets have to be selected carefully. 
Moreover, this experiment shows that our definition of a beneficial grouping set is applicable because (i) not all grouping sets are beneficial and (ii) not all queries can or need to be optimized with a grouping set.
The benefit criterion considers both aspects to reduce workload execution times.

\subsection{Main Findings}
For both types of ML models for cardinality estimation, our \emph{aggre\-ga\-tion-enabled training phase} offers a unique database-centric way to reduce execution time. 
Table~\ref{tab:runtimes} summarizes our evaluation results.
The overhead introduced by the construction of a grouping set is much smaller than the savings in execution time.
So, grouping sets reduce the workload execution times and amortize their own construction time.
Overall, the benefits of grouping sets as a \emph{aggregation-enabled training phase} for machine learning-based cardinality estimation can be shown.
This is advantageous, especially for the local model approach.
The simpler structure of the local model workloads is better supported by grouping sets because they contain fewer combinations of columns and fewer distinct values.
These are exactly two of the assets for grouping sets identified in Section \ref{sec:aggregation}.
This leads to a significantly higher performance speed up for local model workloads than for global model workloads.
Thus, we can afford a larger amount of local models to reach the schema coverage of a global model.
Even if these models request more queries than the global model, their benefits from the use of grouping sets outweigh the higher number of queries.
This makes up for the major disadvantage of the local approach to have numerous models to be competitive to a global model.

\section{Related Work}
\label{sec:related}
In this section, we detail the importance of database support for machine learning in other works.
We look at the motivation for pre-aggregates from both the database system and the machine learning point of view.

When looking at the synergy of database and machine learning systems, there are three possible interactions: (i) integrate machine learning into database systems, (ii) adapt database techniques for machine learning models, and (iii) combine database and machine learning into one life cycle system~\cite{kumar2017data}.
Based on that, we classify our work in category (iii). 
However, the focus in this area is more on feature and model selection and not on sampling example data.
We argue that the direct support of machine learning training phases with databases should be treated with the same attention. 

To best of our knowledge, there is only little research on directly optimizing the sampling of workloads for machine learning problems.
The authors of~\cite{kipf2018learned} detail their method of speeding up the global model query sampling in~\cite{kipf2019deepsketches}.
They use massive parallelism by distributing the workload over several DB instances.
We see this as a promising step because our approach can also profit from parallel execution.
Especially the instantiation and the querying of grouping sets can be done in parallel because grouping sets are orthogonal to each other.
This means each grouping set aggregates different parts of the original data.

Another thing to look at is the availability of supportive data structures in database systems.
The cube operators are established in databases and benefit from a wide-ranged support~\cite{DBLP:conf/vldb/AgarwalADGNRS96,gray1996cube,DBLP:conf/sigmod/HarinarayanRU96,shukla1996storage,DBLP:conf/sigmod/ZhaoDN97}.
The ability of a database to deliver necessary meta information is also important.
For example, fast querying for the distinct values of each column has a large impact on performance.
A simple solution for this is a dictionary encoding of the data in the database.
Some database systems already use dictionary coding for all their data~\cite{faerber2012hana}.
This is beneficial for our approach because from a dictionary encoded column it is easy to yield the number of distinct values with a dictionary scan.
This scan is cheaper than a whole table scan.
Moreover, dictionary encoding directly supports the transformation of the data into the grouping set dimension and the definition of ranges of these dimensions.
These two properties are inferred from the keys and the value ranges of the dictionary mapping.

Aside from machine learning, database support for data mining has already been an important research topic~\cite{DBLP:conf/kdd/AgrawalS96,DBLP:journals/jiis/ChoWC09,DBLP:conf/vldb/HinneburgLH03,DBLP:conf/icde/NetzCFB01,DBLP:conf/sigmod/OrdonezC00}. 
For example, we identified in~\cite{DBLP:conf/vldb/HinneburgLH03} that aggregation in sub-spaces formed by combinations of attributes is a common task in many data mining algorithms. 
Based on that observation, we see a large potential for tighter coupling of databases and mining algorithms.

\section{Conclusion}
\label{sec:conclusion}

We made the case for cardinality estimation as a candidate for database support of machine learning for DBMS.
We detailed an approach for pre-aggregating count information for cardinality estimation workloads.
It uses grouping sets, a well-known database data structure, to reduce the data to be scanned by example queries for cardinality estimation with machine learning models.
This reduces the execution time of a given workload even though we spend extra time to construct the intermediate data structures.

This case has a strong potential to be applied to the other similar machine learning problems, like plan cost modeling or indexing.
We liken the potential to machine learning workloads for any of these machine learning problems in DBMS where information about the data in the DB is aggregated.
These parallels make grouping sets and therefore DB support beneficial for ML for DBMS in general.

\bibliographystyle{ACM-Reference-Format}
\bibliography{sample-base}


\end{document}